\documentclass[aps,prd,twocolumn,groupedaddress,showpacs,floatfix]{revtex4}
\usepackage{amssymb}
\usepackage{amsmath}
\usepackage{graphicx}
\begin{document}
\title{Late cosmology of brane gases with a two-form field}
\author{Antonio Campos}
\affiliation{Institut f\"ur Theoretische Physik,
             Universit\"at Heidelberg,
             Philosophenweg 16,
             69120 Heidelberg,
             Germany}
\date{\today}

\begin{abstract}
We consider the effects of a two-form field on the 
late-time dynamics of brane gas cosmologies. 
Assuming thermal equilibrium of winding states, we find that 
the presence of a form field allows a late stage of expansion 
of the Universe even when the winding degrees of freedom 
decay into a pressureless gas of string loops.
Finally, we suggest to understand the dimensionality of the
Universe not as a result of the thermal equilibrium condition
but rather as a consequence of the symmetries of the geometry.
\end{abstract}
\maketitle

Brane gas cosmology \cite{Alexander:2000xv} is a scenario 
inspired by string theory which gives an appealing 
resolution to the initial singularity problem of the 
standard cosmological model and proposes a dynamical 
explanation for the number of spatial dimensions of the 
Universe.
In this type of string cosmologies the geometry of the early 
Universe is described by a $D+1$-dimensional spacetime with a 
spatial toroidal section and the dynamics is assumed to be
driven by a gas containing all the extended degrees of freedom, 
or Dp-branes \cite{Polchinski:1998}, that the spectrum of a 
particular string theory could admit. 
This scenario is, in fact, a natural extension of the original 
proposal which considers a Universe filled with a gas of
fundamental strings.
\cite{Brandenberger:1989aj,Tseytlin:1992xk,Tseytlin:1992ss}
(see \cite{Bassett:2003ck} for a recent discussion of some 
cosmological implications).

The fundamental string/brane states in the gas are the 
winding modes, the momentum modes, and the oscillatory
modes. 
This spectrum is invariant under the inversion of the radius 
of the torus --a symmetry which is called {\sl T duality} 
and interchanges winding a momentum modes.
In the basic picture the spacetime starts small, namely with 
a typical size of the order of the string length, at the
Hagedorn temperature.
As the Universe grows the energy of winding modes increases
leading to a confining potential that prevents further
expansion.
In the absence of equilibrium these modes cannot decay and the 
Universe starts contracting. 
As the size decreases the energy of the momentum modes becomes 
important.
Then, the contraction phase is stopped and the Universe is
prevented from reaching a zero size.
The role of both types of modes is dual: winding modes stop 
expansion whereas momentum modes stop contraction.
Without a mechanism to get rid of winding modes all the 
spatial dimensions of the Universe would oscillate around
the string length scale.
Nevertheless, if thermal equilibrium is maintained, and an
adiabatic evolution is assumed so that the expansion rate
is smaller than the interaction rate, energy from the 
winding states can be transferred to the rest of the states 
of the spectrum and, thus, the universe can be kept expanding.
By considering the probability of intersection of the worldvolume 
of two Dp-branes, it follows that the annihilation mechanism can 
take place only if the largest dimensionality of the spacetime is 
$2p+2$.
Modes with large $p$ are more massive and, consequently, must
decay earlier than modes with lower $p$.
Thus, the dimensionality of the space could be reduced successively
creating a hierarchy of small dimensions.
Since the last modes to be annihilated are those with $p=1$, there
is no problem in having a large Universe expanding in three spatial 
dimensions.
Different aspects of this scenario has been studied in 
\cite{Easson:2001fy,Easther:2002mi,Brandenberger:2001kj,Watson:2002nx,Boehm:2002bm,Campos:2003gj,Easther:2002qk,Alexander:2002gj,Kaya:2003py,Kaya:2003vj,Biswas:2003cx,Watson:2003gf,Brandenberger:2003ge}

The cosmology of brane gases has two shortcomings. 
First, causality implies that even with an efficient
annihilation of winding states at least one Dp-brane across a 
Hubble volume should have survived. 
As with domain wall defects, this is incompatible with current
experimental data of the fluctuations in the temperature of the 
microwave background radiation.
A solution to this problem, proposed in \cite{Alexander:2000xv} 
and further developed in \cite{Brandenberger:2001kj} (see also 
\cite{Campos:2003gj}), is to invoke a sufficiently long period 
of cosmological loitering which might have allowed the whole 
spatial extent of the Universe to be in causal contact, so that 
the actual absence of branes can be explained by a microphysical 
process.
The second problem with these scenarios is that they describe the 
very early stages of the Universe only, and a mechanism to connect 
this string theory regime to a late classical cosmological 
evolution is still needed.
This is, in fact, connected with the problem of dilaton potential
generation which is still not fully understood and is one of the most
outstanding challenges for string theory.

In previous analysis of string/brane gas cosmology the effects of
fluxes have been ignored and the background dynamics has 
been described by a gravitational field and a dilaton field only. 
The purpose of the present work is to investigated the importance
of the Neveu-Schwarz--Neveu-Schwarz (NS-NS) sector on the evolution 
of this type of cosmologies.
As we will see, the effects of fluxes can dominate over winding or
unwinding degrees of freedom at late times and introduce significant 
modifications on the cosmological evolution of the Universe.
As an immediate implication, we suggest to understand the number
of spatial dimensions of the spacetime as a consequence of homogeneity. 

The cosmological dynamics of this scenario is dictated by  
the bosonic action \cite{Alexander:2000xv},
\begin{equation}
S_B
   = -  \int d^{D+1}x\, \sqrt{-g}\, e^{-2\phi}
        \left[ R 
              +4 (\nabla_{\mu}\phi)^2
              -\frac{1}{12}H^2_{\mu\nu\alpha}
        \right]\, , 
\label{eq:b_action1}
\end{equation}
which involves the dynamics of the metric tensor $g_{\mu\nu}$, a 
dilaton field $\phi$, and a two-form field $B_{\mu\nu}$ through 
its field strength $H_{\mu\nu\alpha}=3\nabla_{[\mu}B_{\nu\alpha ]}$.
(We employ units in which the string length is $l_{st}\sim 1$.)
This low-energy effective action describes a type II superstring 
theory without a Ramond-Ramond sector and arises from the
compactification of 11-dimensional supergravity on a circle.
The dilaton field represents the radius of compactification
and defines the string coupling, which is assumed to be small
in order to ensure an effective compactification.
The equations of motion derived from this action are,
\begin{eqnarray}
R_{\mu\nu} + 2\nabla_\mu\nabla_\nu\phi
& = & 
\frac{1}{4}H_{\mu\alpha\beta}H^{\alpha\beta}{}_\nu
\nonumber\\ 
&   &
-\frac{e^{2\phi}}{\sqrt{-g}}
 \left( \frac{\delta S_m}{\delta g^{\mu\nu}}
       -\frac{1}{4} g_{\mu\nu} \frac{\delta S_m}{\delta \phi}
 \right)\, ,
\nonumber\\
2\Box\phi -4(\nabla_{\mu}\phi)^2
& = & 
-\frac{1}{6} H^2_{\mu\nu\alpha}
\nonumber\\ 
&   &
+\frac{e^{2\phi}}{\sqrt{-g}}
 \left( g^{\mu\nu}\frac{\delta S_m}{\delta g^{\mu\nu}}
       -\frac{D-1}{4} \frac{\delta S_m}{\delta \phi}
 \right)\, ,
\nonumber\\
\nabla^\mu(e^{-2\phi}H_{\mu\nu\alpha})
& = &
0\, .
\nonumber
\end{eqnarray}
The gas of Dp-branes is represented by the matter action $S_m$.
Since we are interested in analysing the late-time cosmological 
evolution, we will assume that all winding states with $p>1$ have 
already decayed and only modes with $p=1$ are still interacting 
in a expanding background with three spatial dimensions.
Here we shall concentrate the analysis on spatially flat 
homogeneous and isotropic metrics,
\begin{equation}
ds^2
   = - N^2(t)dt^2
     + e^{2\lambda(t)}\sum^{3}_{i=1}dx^2_i\, .
\label{eq:metric}
\end{equation}
We have kept the lapse function, $N(t)$, in order to
facilitate a later rederivation of the full set of equations
of motion.
A solution to the above dynamical equations can be obtained by 
adopting the following ansatz for the field strength (see for 
instance
\cite{Freund:1980xh,Freund:1982pg,Tseytlin:1992ye,Goldwirth:1994ha,Lidsey:1999mc}),
\begin{equation}
 H_{\mu\nu\alpha}
   = e^{2\phi} 
     \epsilon_{\mu\nu\alpha\beta} \nabla^\beta h\, ,
\end{equation}
where $\epsilon_{\mu\nu\alpha\beta}$ is the totally 
antisymmetric covariantly conserved volume form.
In addition, the field strength satisfies a closure 
condition, $\nabla_{[\beta} H_{\mu\nu\alpha]}=0$ 
(Bianchi identity), which follows from its definition
in terms of an antisymmetric form field, that yields
a differential equation for the scalar function $h$,
$\nabla_\mu(e^{2\phi}\nabla^\mu h)=0$.
A homogeneous solution to this equation, $h(t)$, can 
be easily obtained and leads to a field strength, 
\begin{equation}
H^2_{\mu\nu\alpha}
   = 6 H^2_o e^{-6\lambda}
   \equiv 12 U(\lambda)\, ,
\label{eq:U}
\end{equation}
where $H_o$ is a constant of integration.
Then, for the metric (\ref{eq:metric}), the bosonic action 
(\ref{eq:b_action1}) takes the following simple form,
\begin{equation}
S_B
   = -\int dt\, e^{-\varphi}
        \left[ \frac{1}{N} ( 3\dot\lambda^2 
                            - \dot\varphi^2
                           )  
              -N U(\lambda)
        \right]\, ,
\label{eq:b_action2}
\end{equation}
where the new "shifted" dilaton field $\varphi\equiv 2\phi-3\lambda$
serves to absorb the space volume factor in the action.
It is important to observe that the net effect of the form 
field is to yield an effective exponential potential for the 
scalar field $\lambda$. 
As we will see later this feature will have important 
consequences on the late cosmological dynamics.

By using the gauge $N(t)=1$, the equations of motion 
derived from the total effective action (bosonic plus 
matter sectors) can be written as 
\cite{Tseytlin:1992xk,Tseytlin:1992ss,Tseytlin:1992ye,Veneziano:1991ek,Gasperini:2002bn},
\begin{eqnarray}
 \dot\varphi^2 - 3\dot\lambda^2
&=& e^\varphi E + U(\lambda)\, ,
\label{eq:contraint}
\\
 \ddot\varphi
-3\dot\lambda^2
&=& \frac{1}{2}\,e^\varphi E\, ,
\label{eq:dyn_varphi}
\\
 \ddot\lambda
-\dot\varphi\dot\lambda
&=& \frac{1}{2}\,e^\varphi P + U(\lambda)\, ,
\label{eq:dyn_lambda}
\end{eqnarray}
where $E$ and $P$ are the total energy and pressure
build on all (winding and non-winding) matter source 
contributions,
\begin{eqnarray}
E  &=& E_w + E_{nw}\, ,
\\
P  &=& P_w + P_{nw}\, .
\end{eqnarray}
The assumption of an adiabatic evolution of matter previously
mentioned implies the conservation equation 
$\dot E + 3\dot\lambda P =0$ \cite{Tseytlin:1992xk}.
Note that this condition, together with Eqs.~(\ref{eq:dyn_varphi}), 
(\ref{eq:dyn_lambda}), and (\ref{eq:U}), ensures
that the energy constraint (\ref{eq:contraint}) is 
a conserved quantity.
For the non-winding modes we consider an ordinary 
barotropic equation of state $P_{nw}=\gamma E_{nw}$ 
with $0\leq \gamma \leq 1$.
This includes, for instance, the production of non-winding modes 
in the form of radiation ($\gamma=1/3$) or in the form of a 
pressureless fluid ($\gamma=0$).
On the other hand, for the collective dynamics of winding 
modes we assume a typical equation of state $P_{w}=-E_{w}/3$
\cite{Alexander:2000xv,Boehm:2002bm,Vilenkin:1994}.
It is important to point out that, in principle, one should 
expect a modification of this relation between the total 
winding energy and momentum when the dynamics of a form field 
is taking into account (simply because the effective action for 
individual branes depends explicitly on the field $B_{\mu\nu}$ 
induced on the worldvolume \cite{Polchinski:1998} (see also 
\cite{Leigh:1989jq})).
However, we shall consider in our analysis that this is
not a dominant effect on the full equation of state for
the brane gas and, consequently, that the mass energy dominates 
over all other mode fluctuations.
In fact, if the form field induced on the brane is viewed as a 
small deviation from the induced metric the corrections 
to the equations of motion will be of quadratic order.
Then, although our assumption presents some limitations, it will
give an accurate description as long as the field strength
does not grow very large.

We shall first ignore the contribution of all matter 
sources and consider the case in which $E=P=0$.
Then, Eqs.~(\ref{eq:contraint})-(\ref{eq:dyn_lambda}) have the 
following general solution,
\begin{equation}
\lambda(\eta)
   =  \frac{\sqrt{3}}{6}(A\eta + B)
     +\frac{1}{2}
      \ln\left[\frac{H_o}{A}\cosh A(\eta-\eta_o)
         \right]\, ,
\end{equation}
\begin{equation}
\varphi(\eta)
   = -\frac{\sqrt{3}}{2}(A\eta + B)
     -\frac{1}{2}
      \ln\left[\frac{H_o}{A}\cosh A(\eta-\eta_o)
         \right]\, ,
\end{equation}
where the time parameter $\eta$ is defined by
$dt=e^{-\varphi(\eta)}d\eta$.
Here, $A$, $B$, and $\eta_o$ are constants of integration.
An expression for cosmic time, $t$, in terms of $\eta$
can be computed analytically,
\begin{eqnarray}
t(\eta)
   & = & \frac{(2H_o)^{\frac{1}{2}}}{(\sqrt{3}-1)A^{\frac{3}{2}}}
         e^{\frac{\sqrt{3}}{2}(A\eta_o+B)}
         e^{\frac{\sqrt{3}-1}{2}A(\eta-\eta_o)} 
\nonumber\\
   &   & \times
         F\left( -\frac{1}{2},
                  \frac{\sqrt{3}-1}{4};
                  \frac{\sqrt{3}+3}{4};
                  -e^{2A(\eta-\eta_o)}
          \right)\, , 
\end{eqnarray}
where $F$ is the hypergeometric function.
At late times one has, 
\begin{equation}
\eta
   \underset{t\rightarrow \infty}{\longrightarrow}
   \frac{2}{(\sqrt{3}+1)A}\ln t\, ,
\end{equation}
and then the universe expands asymptotically with a scale 
factor,
\begin{equation}
e^{\lambda(t)}
   \underset{t\rightarrow \infty}{\longrightarrow}
   t^{\sqrt{3}/3}\, .
\end{equation}
However, the string coupling gets large,
\begin{equation}
g  \equiv
     e^{\phi}
   \underset{t\rightarrow \infty}{\longrightarrow}
     t^{(\sqrt{3}-1)/2}\, ,
\end{equation}
and the theory becomes strongly coupled \cite{Mukherji:1997ta}.

Turning back to the case in which the universe is driven by a 
gas of branes we first need to supplement the field equations 
(\ref{eq:contraint})-(\ref{eq:dyn_lambda}) with an appropriate 
description of the decay of winding modes into states without 
winding number.
For our analysis we will closely follow the discussion of 
\cite{Campos:2003gj}.
The self-interaction of winding modes can be thought as analogous 
to that of a network of long cosmic strings decaying into small 
string loops
\cite{Bennett:1986qt,Bennett:1986zn,Vilenkin:1994,Brandenberger:1994by}
(For a detailed account of brane interactions in an expanding 
universe see also \cite{Majumdar:2003da}).
At time $t$ the total winding energy, $E_w$, and the total energy of 
the string loops (states without winding number) produced, $E_l$, can 
be expressed in general as,
\begin{eqnarray}
E_w(t) 
   & = & \tau l_c N_w(t)e^{\lambda(t)}\, ,
\label{eq:winding_energy}
\\
E_l(t) 
   & = & \tau l_c N_l(t) e^{-3\gamma\lambda(t)}\, ,
\label{eq:loop_energy}  
\end{eqnarray}
where $N_w$ and $N_l$ represent the number of winding modes
and small loops, respectively.
$\tau$ is the tension corresponding to a D1-brane and $l_c$
is a derived length scale related with the initial size of
the spatial dimensions of the torus 
$l_c\sim l_{st}\exp(-\lambda(0))$.
The decrease of winding energy, due to the decay into 
small loops, can be roughly estimated to be proportional to
the square of the number of winding modes \cite{Vilenkin:1994},
\begin{equation}
\dot E_w
   \sim -c\, \tau N^2_w(t) 
             \left( \frac{L_l(t)}{R(t)} 
             \right)\, .   
\label{eq:energy_decay_rate}
\end{equation}
The parameter $c$ measures the efficiency of loop production
and, correspondingly, it is zero if there is no winding mode
annihilation.
$L_l(t)$ represents the typical size of the loops created and
it will be taken proportional to the cosmic time $t$
\cite{Bennett:1986zn} (see also \cite{Brandenberger:2001kj} and
\cite{Campos:2003gj}).
{}From Eq.~(\ref{eq:energy_decay_rate}) and within the adiabatic 
approximation we can find evolution equations for $N_w(t)$ and 
$N_l(t)$,
\begin{eqnarray}
\dot  N_w(t)
   & = & -\frac{c}{l^2_c} L_l(t) N^2_w(t) e^{-2\lambda(t)}\, ,
\label{eq:winding_decay_rate}
\\ 
\dot  N_l(t)
   & = &  \frac{c}{l^2_c} L_l(t) N^2_w(t) e^{(3\gamma-1)\lambda(t)}\, .
\label{eq:loop_decay_rate}
\end{eqnarray}
To keep the discussion simple we do not consider more 
elaborated annihilation processes such as those taking 
into account the reconnection of small loops or a subsequent 
loop decay into gravitational radiation.
Finally, the field equations (\ref{eq:contraint})-(\ref{eq:dyn_lambda}) 
can be transformed into a closed set of first-order 
differential equations by introducing the new variables 
$f=\dot\varphi$ and $l=\dot\lambda$,
\begin{eqnarray}
\dot f
&=& 3l^2
   +\frac{1}{2}\,e^\varphi 
    \left( E_w + E_l
    \right)\, ,
\label{eq:dot_f} \\
\dot l
&=& fl
   -\frac{1}{6}\,e^\varphi 
    \left( E_w -3\gamma E_l
    \right)
   +U(\lambda)\, , 
\label{eq:dot_l} \\
f^2 
&=& 3l^2
    +e^\varphi
     \left( E_w + E_l
     \right)
    +U(\lambda)\, ,
\label{eq:constraint}
\end{eqnarray}
together with Eqs.~(\ref{eq:U}), (\ref{eq:winding_energy}), 
(\ref{eq:loop_energy}), (\ref{eq:winding_decay_rate}), and 
(\ref{eq:loop_decay_rate}).

\begin{figure}[t]
\includegraphics*[totalheight=1.1\columnwidth,width=0.95\columnwidth]{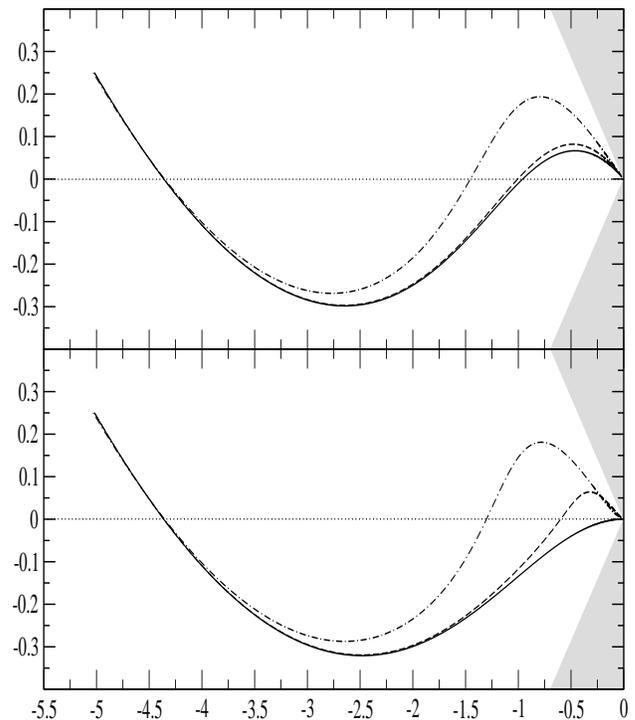}
\caption{Phase space $(f,l)$.
These curves represent the numerical solutions of the equations 
of motion for an efficiency parameter of the winding mode decay 
$c=0.1$, and a gas of small loops with $\gamma=1/3$ (top) and
$\gamma=0$ (bottom).
The straight lines correspond to $H_o=0$, the dashed lines 
to $H_o=0.001$, and the dotted-dashed lines to $H_o=0.005$.
The grey area, $f^2-3l^2<0$, represents a region where the total 
matter energy is negative.
The border lines, $l=\pm f/\sqrt{3}$, are straight solutions
with $E_w=E_l=U=0$ and therefore cannot be crossed by other
trajectories. 
\label{fig:phase_space_EQ}}
\end{figure}

Let us comment on the general properties of the above 
dynamical equations. 
The straight lines $l=\pm f/\sqrt{3}$ are solutions 
with $E_w=E_l=U=0$.
Since $U(\lambda)$ is nonnegative, the region with
$f^2-3l^2<0$ represents configurations with a negative
total matter energy and therefore it is excluded from
the analysis (see Fig.~\ref{fig:phase_space_EQ}). 
On the other hand, Eq.~(\ref{eq:dot_f}) implies that
$\dot f>0$ and then $f$ is always growing.
If initially $f$ is negative it will approach zero 
asymptotically and $\varphi$ will be a decreasing 
function of time.
For those trajectories that obey the condition $f-3l<0$, 
the dilaton is also decreasing and thus one can be certain 
that the weak coupling condition is enforced 
\cite{Campos:2003gj}.
The evolution of the variable $l$ (the Hubble function) 
is dictated by Eq.~(\ref{eq:dot_l}) which has a simple 
mechanical interpretation: it describes the classical 
motion of a damped particle in a time-dependent potential, 
which in our case is given by 
\begin{eqnarray}
\tilde U(\lambda)
   & = & \frac{\tau l_c}{6}\, e^\varphi
         \left[ N_w(t)\, e^\lambda
               +N_l(t)\, e^{-3\gamma\lambda}
         \right] 
        +\frac{1}{6} U(\lambda)
\nonumber\\
   & = & \frac{\tau l_c}{6}\, e^\varphi
         \left[ N_w(t)\, e^\lambda
               +N_l(t)\, e^{-3\gamma\lambda}
               +\frac{H^2_oe^{-\varphi}}{2\tau l_c}\,
                e^{-6\lambda}
         \right].
\nonumber\\
\label{eq:clas_pot}
\end{eqnarray}
The analysis of this effective potential gives a
valuable information of the full dynamics.
First, consider the flux is negligible ($H_o=0$) 
and the gas of branes is out of thermal equilibrium.
Then, $N_l=0$ and $N_w$ is approximately constant. 
This situation leads to an oscillating cosmological 
evolution.
If the Universe is growing the winding term dominates
and tends to stop the expansion.
On the other hand, when it is contracting the momentum
modes become important and  causes the Universe to
reexpand.
(These modes can be roughly represented by a potential 
similar to the second term with $\gamma=1/3$ and $N_l$ a 
non vanishing constant.)
This behaviour, which typically describes the dynamics 
close to the Hagedorn temperature \cite{Brandenberger:1989aj}, 
changes once thermal equilibrium has been established. 
Now winding energy can be transfered to matter 
energy and the second term starts to dominate as $N_l$ 
increases and $N_w$ decreases.
Under these conditions, the Universe can pass through a 
contraction phase, the longer the less efficient the energy 
transfer is, but it shall inevitably expand in the future.
Nevertheless, as it was shown in \cite{Campos:2003gj}, if 
winding modes decay into a pressureless gas ($\gamma=0$) 
of small string loops, (\ref{eq:clas_pot}) always 
represents a confining potential and the Universe cannot 
expand and grow large.
As a consequence, the dimensionality of the spacetime cannot 
be explained by this mechanism in this particular case. 

The importance of the two-form field contribution, on the other 
hand, is more difficult to assess without solving the equations 
of motion, because it also involves the dynamics of the field 
$\varphi$.
Naively, one should expect that the last term in the effective 
potential (\ref{eq:clas_pot}) could dominate the dynamics only 
if $\varphi$ becomes negatively large.
In Fig.~\ref{fig:phase_space_EQ} the numerical solution of the 
full set of dynamical equations has been plotted for different 
parameters of the model.
As one can observe, the contribution from the flux is only 
significant at late times.
When the small loops are produced in the form of a gas with 
$0<\gamma\leq 1/3$ (in Fig.~\ref{fig:phase_space_EQ} the case 
of a gas of radiation, $\gamma=1/3$, has been plotted), the 
cosmological evolution do not change qualitatively. 
Mainly, the effect of the two-form field is to reduce the rate of 
contraction and increase the rate of late expansion.
However, if the loops produced behave as ordinary static matter with
zero pressure, $\gamma=0$, the qualitative change is much more
interesting.
In this case, contrary to what happens when there is no flux field
(the case $H_o=0$ in Fig.~\ref{fig:phase_space_EQ}), the Universe
is able to escape from the contraction phase induced by the winding 
mode potential and finally reexpands.
This behaviour overcomes the obstruction for explaining the 
dimensionality of the spacetime previously mentioned for a 
pressureless gas of string loops.
In \cite{Campos:2003gj}, another mechanism was proposed
to solve this problem by considering the dynamics of a gas
of nonstatic branes.
In this case, the equation of state for the gas depends 
explicitly on a constant characteristic average velocity 
that modifies qualitatively the cosmological role of winding 
modes.
For values of this velocity above certain threshold, the 
winding potential is no longer a confining potential. 
Thus winding modes do not tend to stop expansion and the 
universe cannot enter a phase of contraction.
The problem with this mechanism is that the assumption of a 
time independent average velocity is likely very strong and, 
unfortunately, a clear picture of its time evolution is still 
needed.

\begin{figure}[t]
\includegraphics*[totalheight=0.7\columnwidth,width=0.95\columnwidth]{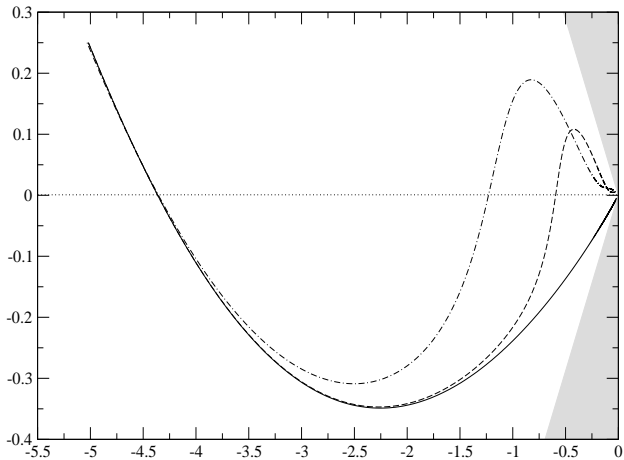}
\caption{Phase space $(f,l)$.
These curves represent the numerical solution of the equations 
of motion without winding mode annihilation ($c=0$).
The straight line correspond to $H_o=0$, the dashed line 
to $H_o=0.001$, and the dotted-dashed line to $H_o=0.005$.
\label{fig:phase_space_noEQ}}
\end{figure}

To sum up, we see that under thermal equilibrium the dynamics 
of a flux field dominates at late times over the nonwinding 
degrees of freedom.
The next obvious question to ask is whether the form field
term in the effective potential also dominates if there is no 
winding mode decay.
This would immediately imply that the confining potential of the
winding states will not be able to stop the expansion of the 
Universe, and consequently, one could question if the 
assumption of thermal equilibrium is essential in order to 
explain the dimensionality of the spacetime.
To investigate this issue we have solved the equations of
motion considering that the winding modes are not 
annihilated by self-interactions.
As can be seen from Fig.~\ref{fig:phase_space_noEQ}
the early dynamics is dominated by the winding modes
that try to halt the Universe but, at some point in the
evolution, the flux field catches up and then triggers a 
late phase of expansion.
Thus, in principle, when fluxes are taking into account 
one can relax the thermal equilibrium condition and still be 
able to have a Universe with three large spatial directions.
Once the Universe gets out of the string regime and enters a
classical evolution, the decay (for instance, in the form of 
radiation) of winding modes, and therefore thermal equilibrium, 
will eventually be required to avoid an undesirable future 
expanding evolution because the energy density of these modes 
would dominate over other forms of matter.

What we would like to stress is the possibility that thermal 
equilibrium could not be a fundamental prerequisite for 
explaining the number of space dimensions of the Universe.
Instead, we suggest that it is the symmetry imposed on the
geometry the assumption which is essential.
The most simple picture one can think of is that of a homogeneous
and isotropic Universe with $D+1$ small spacetime dimensions in a 
initial state consisting of a flux field, and winding and momentum
degrees of freedom not necessarily in thermal equilibrium.
The existence of a spatially homogeneous three-form field strength 
is compatible with a homogeneous and isotropic spacetime only in 
two (electric solution) or three (magnetic solution) spatial 
dimensions \cite{Lidsey:1999mc} which implies that the flux has 
to live in a submanifold of the full spacetime.
One can think of this lower dimensional subspace as a D2- or D3-brane 
carrying some charge.  
For the electric solution the field strength contributes to the 
effective potential (\ref{eq:clas_pot}) with a confining term of the 
form  $e^{6\lambda}$, thus the Universe will always oscillate and 
remain small.
For the magnetic solution, nevertheless, the situation is quite 
different. 
To begin with, consider the Universe is initially in an expansion 
stage driven by momentum modes.
As the Universe grows, the winding energy will start to dominate
and oppose the expansion.
Then, the Universe can start contracting in all its spatial 
directions.
This goes on until the field strength contribution builds up and 
the contraction is reversed in the submanifold containing the flux 
field.
The expansion of this subspace cannot be stopped by winding modes 
and then a 3+1-dimensional Universe can grow large. 
The rest of the spatial directions will continue contracting and 
expanding indefinitely but they will remain small. 
In that sense it is fundamental for the stability of this internal
space that the winding modes are out of thermal equilibrium.  
In this set up, the initial singularity is also avoided due to the 
presence of momentum modes that oppose contraction as the Universe 
becomes small.
Furthermore, it offers an explanation for the precise number of spatial 
dimensions unlike the original proposal \cite{Brandenberger:1989aj} 
which only provides an upper bound.
It is worth emphasing that all these arguments can be easily 
generalised to an anisotropic picture.

In conclusion, we have studied the late cosmological evolution
of a gas of branes by incorporating the dynamics of the NS-NS 
sector.
We have shown that fluxes can dominate the dynamics of the 
Universe at late times and introduce significant effects. 
Assuming thermal equilibrium, these effects can account for a 
late stage of expansion even if the winding modes decay into a 
pressureless gas of string loops.
Finally, we have argued that assuming homogeneity of the
background the presence of a two-form field can easily explain 
the exact number of spatial dimensions of the Universe and the 
stability of the small internal space without requiring the decay 
of winding modes.
This could mean that the fundamental assumption for understanding
the dimensionality of the spacetime is the symmetry of the geometry 
and not the condition of thermal equilibrium of the winding states.

\mbox{}\\

The author acknowledges the support of the Alexander von Humboldt 
Stiftung/Foundation and the Universit\"at Heidelberg.

\bibliography{bgc}

\end{document}